\documentclass[journal]{IEEEtran}
\usepackage{xcolor,soul,framed} 

\colorlet{shadecolor}{yellow}
\usepackage[pdftex]{graphicx}
\graphicspath{{../pdf/}{../jpeg/}}
\DeclareGraphicsExtensions{.pdf,.jpeg,.png}

\usepackage[cmex10]{amsmath}
\usepackage{cite}
\usepackage{tabularx}
\usepackage{flushend}
\usepackage{tablefootnote}
\usepackage{array}
\usepackage{mdwmath}
\usepackage{amsmath}
\usepackage{amsfonts}
\usepackage{amssymb}
\usepackage{mdwtab}
\usepackage{eqparbox}
\usepackage{url}
\usepackage{xcolor}
\usepackage{comment}
\usepackage{footnote}
\usepackage{multirow}
\usepackage{graphicx}
\usepackage{caption}
\usepackage{algorithm, algorithmicx}
\usepackage{algpseudocode}

\usepackage{float}
\usepackage{placeins}

\usepackage{url} 
\usepackage{array, makecell}

\usepackage{tikz}

\usetikzlibrary{fit,positioning}
\usetikzlibrary{bayesnet}
\usetikzlibrary{arrows}

\begin{document}
\bstctlcite{IEEEexample:BSTcontrol}
    \title{Transformer Meets Gated Residual Networks To Enhance PICU's PPG Artifact Detection Informed by Mutual Information Neural Estimation}

    \author{Thanh-Dung Le,~\IEEEmembership{Senior Member,~IEEE,} Clara Macabiau,  Kevin Albert, \\ 
    Symeon Chatzinotas,~\IEEEmembership{Fellow,~IEEE,}
    Philippe Jouvet, and Rita Noumeir,~\IEEEmembership{Member,~IEEE}

 \thanks{This work was supported in part by the Natural Sciences and Engineering Research Council (NSERC), in part by the Institut de Valorisation des données de l’Université de Montréal (IVADO), in part by the Fonds de la recherche en sante du Quebec (FRQS).}

\thanks{Thanh-Dung Le is with the Biomedical Information Processing Lab, Ecole de Technologie Superieure, University of Quebec, Canada, and also is with the Interdisciplinary Centre for Security, Reliability, and Trust (SnT), University of Luxembourg, Luxembourg (Email: thanh-dung.le@uni.lu).}

\thanks{Symeon Chatzinotas is with the Interdisciplinary Centre for Security, Reliability, and Trust (SnT), University of Luxembourg, Luxembourg.} 

\thanks{Kevin Albert, and Philippe Jouvet are with the CHU Sainte-Justine Research Center, University of Montreal, Montr\'{e}al, Qu\'{e}bec, Canada.}

\thanks{Clara Macabiau and Rita Noumeir are with the Biomedical Information Processing Lab, \'{E}cole de Technologie Sup\'{e}rieure, University of Qu\'{e}bec,  Montr\'{e}al, Qu\'{e}bec, Canada.}
 
}

\markboth{IEEE, VOL., NO., 2025.
}{Thanh-Dung Le \MakeLowercase{\textit{et al.}}: Transformer Meets Gated Residual Networks To Enhance PPG Artifact Detection Informed by Mutual Information Neural Estimation}


 \maketitle

\begin{abstract}
This study delves into the effectiveness of various learning methods in improving Transformer models, focusing mainly on the Gated Residual Network (GRN) Transformer in the context of pediatric intensive care units (PICU) with limited data availability. Our findings indicate that Transformers trained via supervised learning are less effective than MLP, CNN, and LSTM networks in such environments. Yet, leveraging unsupervised and self-supervised learning on unannotated data, with subsequent fine-tuning on annotated data, notably enhances Transformer performance, although not to the level of the GRN-Transformer. Central to our research is analyzing different activation functions for the Gated Linear Unit (GLU), a crucial element of the GRN structure. We also employ Mutual Information Neural Estimation (MINE) to evaluate the GRN's contribution.
Additionally, the study examines the effects of integrating GRN within the Transformer's Attention mechanism versus using it as a separate intermediary layer. Our results highlight that GLU with sigmoid activation stands out, achieving 0.98 accuracy, 0.91 precision, 0.96 recall, and 0.94 F1 score. The MINE analysis supports the hypothesis that GRN enhances the mutual information (MI) between the hidden representations and the output. Moreover, using GRN as an intermediate filter layer proves more beneficial than incorporating it within the Attention mechanism. In summary, this research clarifies how GRN boosters GRN-Transformer's performance surpasses other techniques. These findings offer a promising avenue for adopting sophisticated models like Transformers in data-constrained environments, such as PPG artifact detection in PICU settings.

\end{abstract}


\IEEEpeerreviewmaketitle

\section{Introduction}

Recently, the PICU at CHU Sainte-Justine (CHUSJ) has achieved significant progress by establishing a high-resolution research database (HRDB) \cite{brossier2018creating, roumeliotis2018reorganizing}. This state-of-the-art database seamlessly integrates biomedical signals from various monitoring devices into the electronic patient record, enhancing data continuity throughout a patient's PICU stay \cite{mathieu2021validation}. The implementation of HRDB has notably improved the Clinical Decision Support System (CDSS) at CHUSJ, elevating patient safety and supporting decision-making with substantial evidence \cite{dziorny2022clinical}. A critical aim of the CDSS at CHUSJ is the prompt and precise diagnosis of acute respiratory distress syndrome (ARDS). Monitoring Oxygen saturation (SpO2) values, crucial for ARDS diagnosis, is key in predicting and managing ARDS \cite{le2022detecting, sauthier2021estimated}. These values are also vital in determining respiratory support strategies \cite{emeriaud2023executive, jouvet2012pilot, wysocki2014closed}. Additionally, the ability to predict SpO2 from Photoplethysmography (PPG) waveforms and non-invasive blood pressure estimation is increasingly acknowledged as crucial for enhancing CDSS functionalities \cite{hill2021imputation, fan2017estimating}. Therefore, accurately identifying and discarding erroneous waveforms and SpO2 values from CDSS inputs is of utmost importance. Maintaining the accuracy of these inputs is crucial for the operation of the CDSS, thereby directly influencing patient outcomes and the efficiency of care.

In the current landscape of clinical data application, recent researches focus on improving PPG artifact detection through advanced machine learning (ML) techniques to enhance diagnostic accuracy in settings like the PICU. While previous studies, including those by Macabiau et al. \cite{macabiau2023label}, have explored ML approaches in PPG artifact detection, they highlight ongoing challenges such as limited data availability and significant class imbalances, which restrict the performance of fully supervised ML models. Traditional Transformer models, while promising due to their powerful attention mechanisms, have proven less effective in such constrained scenarios. Compared to these models, semi-supervised methods like label propagation and traditional algorithms such as K-Nearest Neighbors (KNN) have provided reasonable accuracy but still fall short in environments with high variance and limited annotations.

Another work introduces a novel architecture to address these challenges by integrating GRN into the Transformer model, creating the GRN-Transformer hybrid \cite{le2023grn}. This model seeks to leverage GRN's structural advantages in handling limited data and class imbalances, aiming to improve artifact detection accuracy and reliability over traditional approaches. However, the performance of supervised approaches remains tied to the availability of annotated data, a well-known limitation in clinical environments \cite{le2024boosting}.

In response to this constraint, an alternative approach explored self-supervised learning (SSL) as an alternative to supervised training to enhance the Transformer’s performance under data-limited conditions \cite{le2024boosting}. Although SSL helped improve the model’s robustness, it did not match the performance gains observed with the GRN-Transformer regarding artifact detection accuracy. Therefore, as shown in Table \ref{tab:transf_compa}, a comparative analysis of supervised, unsupervised, and self-supervised methods reveals the GRN-Transformer as the most proficient model, particularly in its accuracy and recall performance, underscoring the value of the GRN-Transformer's integration.

The motivation behind this study extends further into understanding the role of the GLU within the GRN structure. Studies have shown GLU’s effectiveness in various NLP tasks, often attributed to its ability to optimize perplexity scores and enhance language understanding capabilities \cite{shazeer2020glu}. In the context of the GRN-Transformer, GLU serves as a core component, yet its specific impact on artifact detection within our framework remains largely unexplored. First, we seek to clarify this role by examining different GLU activation functions to determine the most effective configurations for handling limited clinical data. Then, to quantify the GRN’s contribution to the model’s performance more rigorously, we employ MINE \cite{belghazi2018mutual}. MINE enables us to quantify and analyze the flow of information between the GRN and the Transformer components, revealing how effectively the GRN captures and transfers information across layers. By facilitating reliable MI estimation, MINE aids in identifying dependencies that contribute to performance improvements. These insights are valuable for understanding the GRN’s role and optimizing the overall network architecture for tasks that require handling PPG artifact detection.

Despite the growing availability of high-resolution PICU waveforms, reliable PPG-artifact detection remains elusive because (i) labelled data are scarce and highly imbalanced, and (ii) standard Transformer blocks overfit or under-utilise such limited information. Our objective is therefore two-fold: (1) to determine whether a Transformer augmented with a GRN and carefully chosen gated activations can overcome these data constraints, and (2) to quantify, through MI analysis, the extent to which the GRN improves representation quality and downstream diagnostic performance. Our explicit contributions include:

\begin{itemize}
    \item \textbf{Systematic activation analysis}: testing 11 gated or smooth activations inside the GRN to identify functions best suited to noisy, low-volume clinical data.
    \item \textbf{GRN-Transformer architecture}: embedding a GRN layer - acting as an intermediate representation filter that down-weights noisy features and amplifies the most predictive patterns, into both standard Attention and lightweight Gated Attention Unit (GAU) stacks \cite{hua2022transformer}, yielding a family of models that trade off accuracy and compute.
    \item \textbf{Information-theoretic explainability}: by applying MINE and t-SNE, we show that the GRN increases feature–label MI by x2.6 and produces cleaner class clusters, directly linking architectural changes to performance gains.
    \item \textbf{Benchmarking on PICU data}: using CHUSJ’s high-resolution database, we demonstrate that the proposed GRN-Transformer outperforms supervised, semi-supervised, and self-supervised baselines while maintaining bedside-friendly compute requirements.
\end{itemize}

The remainder of the manuscript is organised as follows. Section \ref{sec:grn} introduces the GRN-Transformer architecture, and Section \ref{sec:data_CHUSJ} details the CHUSJ dataset and the artifact-annotation protocol. Section \ref{sec:ml_classifiers} discusses the activation-selection methodology, the training procedure, and the MINE-based explainability pipeline. Section \ref{sec:result_discussions} presents quantitative results, MI analyses, and t-SNE visualisations. Section \ref{sec:clinical_implications} discusses clinical implications and outlines deployment considerations. Section \ref{sec:conclusion} concludes and sketches future directions for extending GRN-enhanced Transformers to other vital-sign modalities and low-resource healthcare settings.

\begin{table}[!t]
\centering
\caption{A summary of Transformer's performance for PPG artifact detection at CHUSJ.}
\footnotesize
\label{tab:transf_compa}
\begin{tabular}{|l|c|c|c|c|}
\hline
\multicolumn{1}{|c|}{Transformer-based   Models} & Acc  $(\uparrow)$          & Pre $(\uparrow)$  & Rec  $(\uparrow)$         & F1 $(\uparrow)$            \\ \hline
Supervised \cite{macabiau2023label}                                      & 0.95          & 0.85 & 0.86          & 0.85          \\ \hline
Unsupervised   (AE) \cite{le2024boosting}                            & 0.97          & 0.89 & 0.93          & 0.91          \\ \hline
Self-supervised \cite{le2024boosting} & 0.97 & \textbf{0.93} & 0.92 & \textbf{0.93} \\ \hline
GRN-Transformer \cite{le2023grn}                                 & \textbf{0.98} & 0.90 & \textbf{0.97} & \textbf{0.93} \\ \hline
\end{tabular}
\vspace{1mm} 
\captionsetup{font=scriptsize, justification=raggedright, singlelinecheck=false} 
\caption*{\hspace{0.2mm}\textbf{Bold} denotes the best values.}
\vspace{-2mm}
\end{table}


\section{Gated Residual Network (GRN)}
\label{sec:grn}

\begin{figure}[!tp]
	\centering
	\includegraphics[scale=0.425]{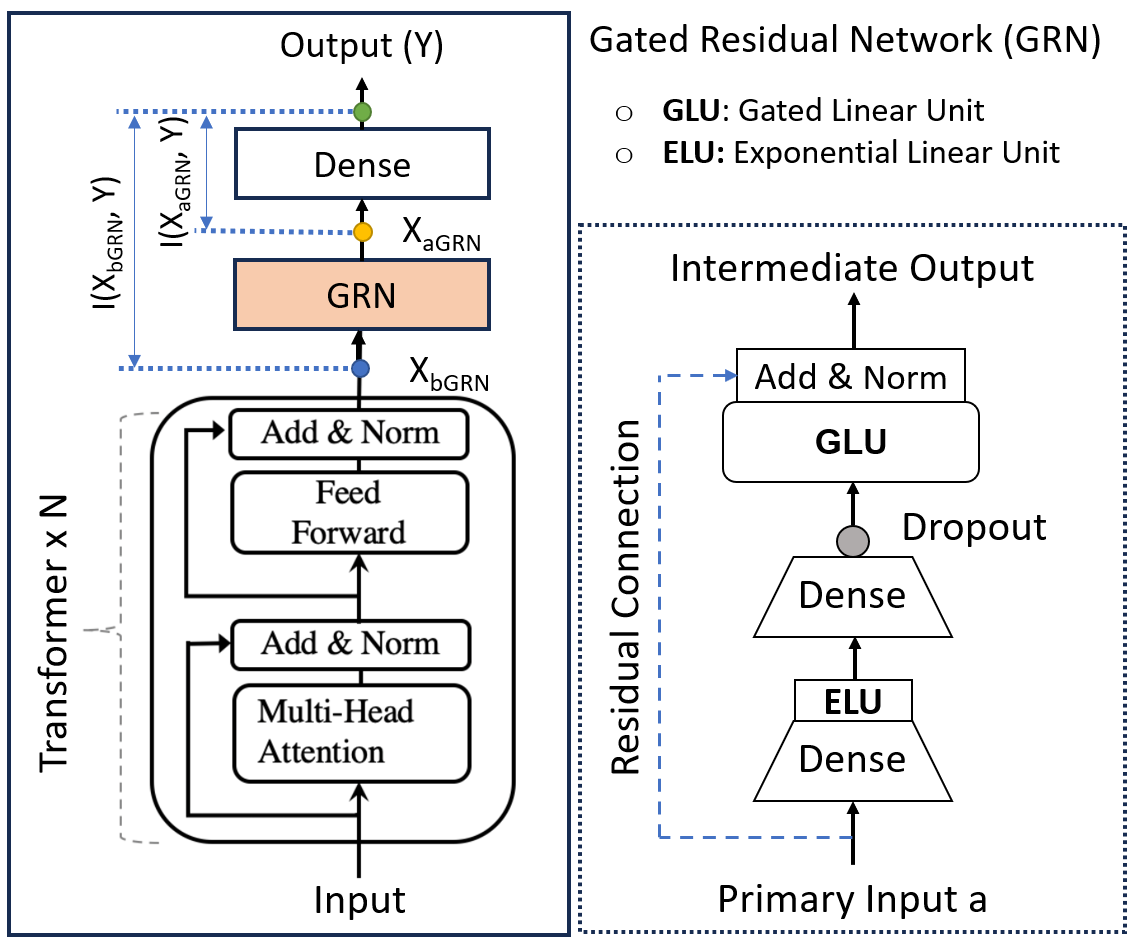}
	\caption{An end-to-end process diagram workflow demonstration. \textit{Left}: an N-layer Transformer encoder processes the input sequence, after which a GRN is inserted as an intermediate representation layer that filters and re-weights the encoder features before the final dense output. \textit{Right}: the GRN itself couples an ELU-activated dense projection with a GLU gate inside a Residual-Add-\&-Norm block; this design adaptively amplifies informative patterns while suppressing noise, providing a cleaner feature space for the downstream classifier.}
	\label{fig:GRN_mine}
        \vspace{-5mm}
\end{figure}

Training Transformer models with small datasets pose significant challenges. These models typically exhibit a generalization gap and tend toward sharp minima in such contexts \cite{le2023small}. Furthermore, their efficacy diminishes when dealing with imbalanced and small PPG signals \cite{macabiau2023label}.

To mitigate these issues, various strategies have been proposed. One approach involves altering the attention mechanism and employing data augmentation methods \cite{lee2021vision}. Alternatively, integrating Convolutional Neural Networks (CNNs) with the Transformer's attention mechanism has been explored \cite{shao2022transformers}. However, these solutions are not without drawbacks:

\begin{enumerate}
\item \textbf{Computational Complexity \cite{hahn2020theoretical}}: Transformers are inherently resource-intensive, with the self-attention mechanism's computational demands scaling quadratically with input length. Adding CNNs can further increase these demands, particularly for lengthy data sequences, potentially making it unfeasible in certain scenarios.
\item \textbf{Sequential Processing in CNNs \cite{sattler2019understanding}}: CNNs process data sequentially, focusing on small, localized regions. This approach hinders their ability to capture long-range dependencies effectively.
\end{enumerate}

In addition to these methods, study \cite{le2023grn} introduces the GRN as a key component of a Transformer-based classifier. Termed the GRN-Transformer, this integration handles small datasets and ambiguous input-target relationships. The GRN effectively handles uncertain input-target relationships, enabling nonlinear processing when necessary. A crucial feature of our GRN is the use of GLU \cite{dauphin2017language}, which dynamically emphasizes or suppresses information based on task requirements. Such gating techniques have been utilized in various models, including Gated Transformer Networks \cite{liu2021gated} and Temporal Fusion Transformers \cite{lim2021temporal}. Its benefits are not limited to time-series data \cite{lim2021temporal, liu2021gated} but extend to a wide range of data types \cite{savarese2017residual, dauphin2017language}. Incorporating GRN into the Transformer architecture marks a significant innovation of our work, greatly enhancing model performance and generalization across various domains.

According to \cite{lim2021temporal}, the GRN processes an input $a$, as illustrated in Fig. \ref{fig:GRN_mine}, with the following output:

\vspace{-3mm}

\begin{align}
\text{GRN}\omega(a) &= \text{LayerNorm}(a + \text{GLU}\omega(\theta_1) ), \\
\theta_1 &= W_{1, \omega}~\theta_2 + b_{1, \omega}, \\
\theta_2 &= \text{ELU}( W_{2, \omega}~a + b_{2, \omega})
\end{align}

Here, $\theta_1$ and $\theta_2$ represent intermediate layers, $\text{LayerNorm}$ denotes standard layer normalization, and $\omega$ indicates shared weights. The Exponential Linear Unit (ELU) activation function, defined as follows for $0 < \alpha$, is also employed:
\begin{align}
f(x)= \begin{cases}x & \text { if } x>0 \\ \alpha(\exp (x)-1) & \text { if } x \leq 0\end{cases}
\end{align}

Additionally, the GRN uses a GLU in its gating layers for architectural flexibility. Given input $\eta$, the GLU operates as follows:
\vspace{-3mm}

\begin{align}
\text{GLU}_{\omega(\eta)} & = \sigma(W_{3, \omega}\eta + b_{3, \omega}) \odot (W_{4, \omega}\eta + b_{4, \omega} ),
\label{eqn:component_gate}
\end{align}

\noindent where $W_{(.)}$ and $b_{(.)}$ represent the weights and biases, respectively, $\odot$ signifies the element-wise Hadamard product, and $\sigma(.)$ denotes the sigmoid activation function:
\vspace{-3mm}

\begin{align}
\sigma(x) = \frac{1}{1 + \exp(-x)}.
\end{align}

The GLU enables the GRN to regulate the extent of its contribution to the input $a$. It can effectively bypass the layer by setting the GLU outputs near zero, thus suppressing non-linear contributions. During training, dropout is applied before the gating layer and layer normalization, specifically to $\theta_1$ in Eq. (2). It has proven that this GRN enhances model robustness and helps prevent overfitting \cite{savarese2017residual}.

As established in prior research, the GLU has demonstrated a critical role in achieving lower perplexities in de-noising tasks. When integrated into Transformer architectures, it has consistently enhanced performance across multiple downstream language understanding applications. Nevertheless, the mechanisms by GLU contribute to these improvements remain insufficiently understood. As noted in \cite{shazeer2020glu}, ``\textit{We offer no explanation as to why these architectures seem to work; we attribute their success, as all else, to divine benevolence.}" This study seeks to address this gap by investigating two central aspects. \textbf{Firstly}, we hypothesize that the activation function is pivotal. In neural networks, the selection of activation functions directly impacts the learning process's efficiency and effectiveness \cite{RamachandranZL18}. Our work will experimentally evaluate the influence of the GLU activation function on Transformer architecture performance and examine how it affects associated learning dynamics. \textbf{Secondly}, we propose leveraging MINE to quantify and assess information flow between the GRN and the Transformer layers. By applying MINE, we aim to determine how effectively the GRN captures and conveys information across layers, shedding light on the dependency structures that underpin the model's performance gains. We anticipate that these insights will elucidate the role of GRN in the network and inform optimization strategies for applications requiring sophisticated dependency, such as PPG artifact detection.
\vspace{-1mm}

\section{Materials and Methods}
\subsection{Clinical PPG Data at CHUSJ}
\label{sec:data_CHUSJ}

The CHUSJ-PICU has established a HRDB that links biomedical signals from patient monitors to electronic patient records, enabling comprehensive physiological data capture through invasive and non-invasive monitoring techniques. This extensive dataset includes measurements from pulse oximetry, providing PPG signals and blood pressure readings obtained via various methodologies. This study, approved by the ethics board of CHUSJ, University of Montreal (approval number eNIMP:2023-4556), focuses on pediatric patients aged 0 to 18 admitted between September 2018 and September 2023. It incorporates electrocardiogram (ECG), PPG, and arterial blood pressure (ABP) waveforms, with data exclusions applied for recordings beyond the fourth day of hospitalization and patients undergoing Extracorporeal Membrane Oxygenation (ECMO) or experiencing multiple admissions. From this selection criteria, data from 1,573 patients were retained, comprising continuous 96-hour ECG, PPG, and blood pressure recordings. These were recorded at 5-second intervals, with a sampling rate of 128 Hz for PPG and 512 Hz for both blood pressure and ECG, and a focused 30-second window of PPG signals was explicitly extracted for analysis.

\begin{figure}[!htp]
	\centering
	\includegraphics[scale=0.65]{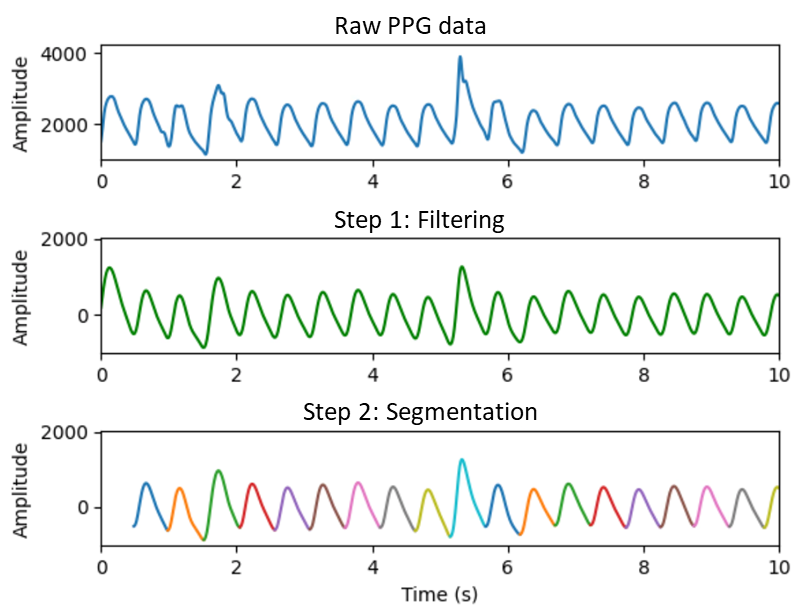}
	\caption{An example for the first two steps of the preprocessing from a 10-second raw PPG signal (top), corresponding filtered signal (middle), and segmented signal (bottom) \cite{macabiau2023label}.}
	\label{fig:preprocessing}
\end{figure}

Data preprocessing in this study consists of four main stages: filtering, segmentation, resampling and normalization, and feature extraction. Initially, a bandpass filter is applied to each signal using a Butterworth filter with cut-off frequencies set between 0.5 Hz and 5 Hz, implemented with a forward-backward approach to preserve signal integrity while reducing noise. The segmentation step follows, whereby the PPG signal is divided into individual pulse segments by identifying local minima, facilitating in-depth artifact analysis within each pulse. Each pulse segment is then uniformly oversampled to 256 samples, representing a 1-second heart cycle, with linear interpolation employed to maintain consistent sampling across pulses. Normalization is subsequently applied to achieve uniform feature scaling. Finally, temporal features are extracted at 4 millisecond intervals within each segment, resulting in 256 samples per pulse. This process effectively captures the temporal dynamics of the PPG signal, preparing it for subsequent analysis and classification using ML methods from traditional to advanced architectures.

The study's data annotation process begins with a healthcare professional manually annotating PPG signal pulses to establish a reliable ground truth, which is crucial for evaluating classification algorithms and classifier performance. To augment this, an automated algorithm, acting as a pseudo-expert, reannotates 10\% of the data initially reviewed by the human expert, using statistical techniques to ensure the pulses fall within expected parameters \cite{macabiau2023label}. This dual-annotation approach enhances the accuracy and reliability of the motion artifact annotations. Furthermore, to optimize the ML algorithms for automatic artifact classification. This approach, involving 1,571 signals with over 81,000 pulses and 256 features per pulse, helps identify the most efficient subset size for annotation. The statistical characteristics of this data, including key metrics such as signal distribution and feature variability, are summarized in Table \ref{tab:summary_statistics}. For a comprehensive overview of the data preprocessing methods employed in this study, we direct the reader to the following studies \cite{macabiau2023label, le2024boosting, le2023grn}.

\begin{table}[ht]
\centering
\caption{Statistical summary of the dataset.}
\footnotesize
\begin{tabular}{|l|c|c|c|}
\hline
Statistic                & Overall & Non-artifact & Artifact\\ \hline
Count                             & 8190             & 6753             & 1437             \\ \hline
Mean                              & 13.53            & 14.98            & 6.70             \\ \hline
Standard Deviation                & 329.36           & 285.95           & 439.81           \\ \hline
Minimum                           & -1784.64         & -1590.12         & -1686.26         \\ \hline
25th Percentile                   & -185.99          & -165.09          & -254.19          \\ \hline
50th Percentile (Median)          & -5.05            & -1.61            & -12.73           \\ \hline
75th Percentile                   & 203.48           & 180.45           & 279.35           \\ \hline
Maximum                           & 2016.81          & 1644.23          & 1981.25          \\ \hline
Skewness                          & 0.23             & 0.39             & -0.00            \\ \hline
Kurtosis                          & 3.04             & 3.28             & 1.70             \\ \hline
\end{tabular}
\label{tab:summary_statistics}
\end{table}

\subsection{Gated Linear Unit with Different Activation Functions}
\label{sec:ml_classifiers}

\begin{table*}[!htp]
\footnotesize
\centering
\caption{List of activation functions \cite{rasamoelina2020review, lederer2021activation}}
\begin{tabular}{ll}
\hline
Name         & Equation                                \\ \hline
Sigmoid      & $\frac{1}{1+e^{-x}}$                    \\
Hard Sigmoid (hard\_$\sigma$)  & $\max\left(0, \min\left(1,\frac{\left(x+1\right)}{2}\right)\right)$      \\
Soft-Sign Sigmoid (SoftSign) & $x / (1 + |x|)$ \\
Snake Periodic Function (Snake) & $x+\sin^2(ax)/a$ \\
Linearly Scaled Hyperbolic Tangent (LiSHT) &$x.\tanh(x)$ \\
Rectified Linear Unit (ReLU)         & $\max(0, x)$                            \\
Exponential Linear Unit (ELU)          & $\max(0, x) + \min(0, \alpha(e^{x} - 1))$ \\
Gaussian Error Linear Unit (GELU) & $x \mathcal{P}(X \leq x), X\sim \mathcal{N}(0,1)$ \\
Scaled Exponential Linear Unit (SELU)         & $\gamma(\max(0, x) + \min(0, \alpha(e^{x} - 1)))$ \\
Sigmoid-Weighted Linear Units (Swish$\beta$)        & $\frac{x}{1+e^{-\beta x}}$                    \\
Self Regularized Non-Monotonic Neural (Mish)         & $x \tanh(\log(1 + e^{x}))$              \\ \hline
\end{tabular}
\label{tab:activation_list}
\end{table*}

\begin{table*}[!htp]
\centering
\footnotesize
\caption{GLU functions with different activations }
\label{tab:activation_glu_functions}
\begin{tabular}{lll}
\hline
\textbf{Models} & \textbf{Activation functions} & \textbf{GLU form functions} \\ \hline
BilinearGLU        & Linear & $(xW + b) \odot (xV + c)$    \\
GLU             & Sigmoid & $\sigma(xW + b) \odot (xV + c)$ \\
hardGLU         & hard\_$\sigma$  & hard\_$\sigma(xW + b) \odot (xV + c)$ \\
SoftsignGLU     & SoftSign   & Softsign$(xW + b) \odot (xV + c)$ \\
SnakeGLU        & Snake  & Snake$(xW + b) \odot (xV + c)$ \\
LiGLU           & LiSHT & LiSHT$(xW + b) \odot (xV + c)$ \\
ReGLU           & ReLU & max$(0, xW + b) \odot (xV + c)$ \\
EGLU            & ELU  & ELU$(xW + b) \odot (xV + c)$ \\
GEGLU           & GELU & GELU$(xW + b) \odot (xV + c)$ \\
SeGLU           & SELU & SELU$(xW + b) \odot (xV + c)$ \\
SwiGLU          & Swish$\beta$ & Swish$\beta$$(xW + b) \odot (xV + c)$ \\
MiGLU           & Mish & Mish$(xW + b) \odot (xV + c)$ \\
 \hline
\end{tabular}
\end{table*}

\begin{table*}[!htp]
\centering
\footnotesize
\caption{Gated Non-Linear Unit (GnLU) functions with different activations }
\label{tab:activation_gnlu_functions}
\begin{tabular}{lll}
\hline
\textbf{Models} & \textbf{Activation functions} & \textbf{GnLU functions} \\ \hline
GnLU             & Sigmoid & $\sigma(xW + b) \odot \sigma(xV + c)$ \\
LiGnLU           & LiSHT & LiSHT$(xW + b)$ $\odot$ LiSHT$(xV + c)$ \\
MiGnLU           & Mish & Mish$(xW + b)$ $\odot$ Mish$(xV + c)$ \\
SeGnLU           & SELU & SELU$(xW + b)$ $\odot$ SELU$(xV + c)$ \\
SwiGnLU          & Swish$\beta$ & Swish$\beta$$(xW + b)$ $\odot$ Swish$\beta$$(xV + c)$ \\
\hline
\end{tabular}
\vspace{-3mm}
\end{table*}

As confirmed, GLU is vital in producing better perplexities for the de-noising objective and better results on downstream language-understanding tasks from Transformer. However, it is still unclear how GLU helps the Transformer in such an experiment \cite{shazeer2020glu}. One to experimentally explain this is the activation function. In neural networks, the choice of activation functions plays a pivotal role in determining the effectiveness and efficiency of learning algorithms \cite{RamachandranZL18}. Table \ref{tab:activation_list} summarizes various activation functions, each with its unique equation and characteristics, essential for different neural network applications. The classic Sigmoid function, known for its smooth gradient, is represented alongside its variant, the Hard Sigmoid, which offers a computationally simpler alternative. The Soft-Sign Sigmoid provides a balanced approach, while the Snake Periodic Function introduces a periodic component to the activation. The Linearly Scaled Hyperbolic Tangent (LiSHT) enhances the traditional tanh function with linear scaling. Widely used in deep learning, the Rectified Linear Unit (ReLU) and its variations like the Exponential Linear Unit (ELU), Gaussian Error Linear Unit (GELU), and Scaled Exponential Linear Unit (SELU) offer different approaches to handling negative input values. The Sigmoid-Weighted Linear Units (Swish) and Self Regularized Non-Monotonic Neural (Mish) functions further extend the repertoire of activation functions, providing flexibility and adaptability in neural network design and performance optimization \cite{rasamoelina2020review, lederer2021activation}.

Table \ref{tab:activation_glu_functions} compares various GLU models, each paired with a distinct activation function to create specialized GLU form functions. The BilinearGLU model employs a linear activation function, resulting in a straightforward GLU form. The standard GLU model uses the sigmoid function, while the hardGLU adapts the hard sigmoid (hard\_$\sigma$) for its gating mechanism. The SoftsignGLU integrates the SoftSign activation, and the SnakeGLU incorporates the periodicity of the Snake function. LiGLU utilizes the Linearly Scaled Hyperbolic Tangent (LiSHT), adding a non-linear, scaled twist. The ReGLU, EGLU, GEGLU, and SeGLU models apply the ReLU, ELU, GELU, and SELU functions, each adding unique characteristics to the gating process. SwiGLU and MiGLU explore the dynamics of Sigmoid-Weighted Linear Units (Swish$\beta$) and the Self Regularized Non-Monotonic Neural (Mish) functions, respectively. Each model's GLU form function follows a similar pattern, combining the chosen activation function with linear transformations to modulate the input signal effectively.

To further investigate non-linear transformations, Table \ref{tab:activation_gnlu_functions} presents various Gated Non-Linear Unit (GnLU) models. Like the GLU forms in Table \ref{tab:activation_glu_functions}, each GnLU variant combines two identical activation functions to apply a non-linear transformation, effectively modulating the input signal. The standard GnLU model utilizes the sigmoid function, while LiGnLU, MiGnLU, SeGnLU, and SwiGnLU models apply LiSHT, Mish, SELU, and Swish$\beta$ functions, respectively. This approach aims to enhance model expressiveness by employing these activation functions in a symmetric gating mechanism.

\subsection{Mutual Information Neural Estimation}
\label{sec:mine}

Following our initial experiments with various activation functions, we employ MINE \cite{belghazi2018mutual} as the next step to further investigate the GRN's role within the Transformer model. MINE provides a robust method for estimating MI between high-dimensional continuous variables using neural networks trained via gradient descent. Its scalability concerning both dimensionality and sample size, as well as its compatibility with back-propagation, makes MINE particularly well-suited for complex neural network applications. Several studies have applied MINE in neural networks to enhance interpretability and performance. For example, \cite{HjelmFLGBTB19} utilized MINE to maximize MI for representation learning, improving feature robustness in neural network embeddings. Similarly, \cite{PooleOOAT19} applied MINE to analyze neural network layers, demonstrating its utility in evaluating MI for better network interpretability.

The motivation for using MINE lies in its capacity to quantify relationships within complex architectures; specifically, in explaining how the GRN functions within a neural network, MINE provides insights into how information flows and is retained or lost across layers. By measuring MI, MINE elucidates the GRN's effectiveness in capturing and transferring information, offering a deeper understanding of its contribution to the Transformer's overall performance. This analysis reveals the GRN’s role in network performance, aiding our understanding of its integration into the Transformer.

MINE estimates MI between high-dimensional continuous random variables using a neural network trained with gradient ascent. The process begins with initializing the neural network's parameters, denoted as \( \theta \), which govern the function \( T_{\theta}(x, z) \) that MINE uses to approximate MI. The estimation procedure iteratively computes a lower bound on the MI by leveraging samples from the joint distribution \( \mathbb{P}_{XZ} \) and the marginal distribution \( \mathbb{P}_{Z} \). In each iteration, a minibatch of \( b \) paired samples \( (x^{(i)}, z^{(i)}) \) is drawn from \( \mathbb{P}_{XZ} \), representing the joint distribution of variables \( X \) and \( Z \). Additionally, \( b \) independent samples \( \tilde{z}^{(i)} \) are drawn from the marginal distribution \( \mathbb{P}_{Z} \).

To estimate the MI, MINE computes the empirical lower bound \( V(\theta) \) as follows:

\begin{equation}
    V(\theta) = \frac{1}{b} \sum_{i=1}^b T_{\theta}(x^{(i)}, z^{(i)}) - \log\left(\frac{1}{b} \sum_{i=1}^b e^{T_{\theta}(x^{(i)}, \tilde{z}^{(i)})}\right)
\end{equation}

This equation represents a lower bound on the MI between \( X \) and \( Z \), where \( T_{\theta}(x, z) \) is a neural network parameterized by \( \theta \) that estimates dependencies between the variables. The first term, \( \frac{1}{b} \sum_{i=1}^b T_{\theta}(x^{(i)}, z^{(i)}) \), estimates the expectation over the joint distribution \( \mathbb{P}_{XZ} \). In contrast, the second term approximates the expectation over the product of the marginals \( \mathbb{P}_{X} \otimes \mathbb{P}_{Z} \). The gradient \( G(\theta) \) of the lower bound \( V(\theta) \) is then calculated with respect to \( \theta \) to obtain a bias-corrected estimate:

\begin{equation}
    G(\theta) = \nabla_{\theta} V(\theta)
\end{equation}

This gradient is used to update the network parameters through gradient ascent:

\begin{equation}
    \theta \leftarrow \theta + \alpha G(\theta)
\end{equation}

where \( \alpha \) represents the learning rate. This iterative process continues until convergence, yielding a flexible estimator for MI; the pseudo-code is summarized in Algorithm \ref{algo:mine}.

To evaluate the impact of the GRN on MI within the Transformer architecture, we implement MINE following the critical steps summarized in Algorithm \ref{algo:mine_grn}. This process measures the MI before and after the GRN is applied, offering insights into how effectively the GRN contributes to information retention and representation quality. By comparing MI values, we can assess the GRN's role in enhancing the sequential processing capabilities of the Transformer.

To compute MI using MINE, we model the dependencies between encoded representations and target labels in the GRN-Transformer framework. Let \( \textbf{before\_grn} \) and \( \textbf{after\_grn} \) represent encoded representations before and after the Gated Residual Network (GRN) is applied, with \( \textbf{y} \) denoting the target labels. The joint distribution \( \mathbb{P}_{XY} \) represents pairs \( (X, Y) \) where \( X \) is the encoded data and \( Y \) is the target. To break dependencies, we also define the marginal distribution \( \mathbb{P}_{X} \otimes \mathbb{P}_{Y} \), which pairs each encoded representation \( X \) with a shuffled target label \( Y \).

MINE estimates MI by leveraging the Donsker-Varadhan representation of the KL-divergence, providing a lower bound on the mutual information \( I(X; Y) \) between \( X \) and \( Y \):
\vspace{-3mm}

\begin{equation}
    I(X; Y) \geq \mathbb{E}_{\mathbb{P}_{XY}}[T_{\theta}(X, Y)] - \log \mathbb{E}_{\mathbb{P}_{X} \otimes \mathbb{P}_{Y}} \left[ e^{T_{\theta}(X, Y)} \right]
\end{equation}

where \( T_{\theta}(X, Y) \) is a neural network parameterized by \( \theta \) that learns to distinguish between samples from the joint distribution \( \mathbb{P}_{XY} \) and the marginal distribution \( \mathbb{P}_{X} \otimes \mathbb{P}_{Y} \). For each minibatch, MINE estimates the joint and marginal predictions. The joint prediction, \( \textbf{joint\_pred} \), is computed by applying \( T_{\theta} \) to pairs \( (X, Y) \) sampled from \( \mathbb{P}_{XY} \), while the marginal prediction, \( \textbf{marginal\_pred} \), is obtained by pairing \( X \) with shuffled \( Y \) to simulate the marginal distribution \( \mathbb{P}_{X} \otimes \mathbb{P}_{Y} \).

The empirical mutual information \( \hat{I}(X; Y) \) for each minibatch is computed as:

\vspace{-3mm}

\begin{align}
    \hat{I}(X; Y) &= \frac{1}{b} \sum_{i=1}^b T_{\theta}(x^{(i)}, y^{(i)}) \nonumber \\ 
    &- \log \left( \frac{1}{b} \sum_{i=1}^b \exp(T_{\theta}(x^{(i)}, \tilde{y}^{(i)})) \right)
\end{align}

where \( b \) is the minibatch size, \( x^{(i)} \) and \( y^{(i)} \) are samples from the joint distribution, and \( \tilde{y}^{(i)} \) is a shuffled sample used to approximate the marginal distribution. The first term, \( \frac{1}{b} \sum_{i=1}^b T_{\theta}(x^{(i)}, y^{(i)}) \), estimates the expectation over the joint distribution \( \mathbb{P}_{XY} \), while the second term, \( \log \left( \frac{1}{b} \sum_{i=1}^b \exp(T_{\theta}(x^{(i)}, \tilde{y}^{(i)})) \right) \), approximates the expectation over the marginal distribution \( \mathbb{P}_{X} \otimes \mathbb{P}_{Y} \).

To assess the GRN's impact on MI, we compute \( \hat{I}_{\text{before}} = \hat{I}(X; Y) \) using \( \textbf{before\_grn} \) and \( \textbf{y} \), and \( \hat{I}_{\text{after}} = \hat{I}(X; Y) \) using \( \textbf{after\_grn} \) and \( \textbf{y} \). By comparing \( \hat{I}_{\text{before}} \) and \( \hat{I}_{\text{after}} \), we can quantify the GRN’s influence on information retention and its contribution to representation quality in the Transformer.

\begin{table*}[!ht]
  \centering
  \caption{Comparison of state-of-the-art and the proposed framework, detailing network architecture (Attention, GAU, and State Space Model (SSM)), activation functions, explainability methods, application domain, and key implementation insights.}
  \label{tab:sota_studies}
  \footnotesize
  \begin{tabular}{
      p{2.0cm}
       p{2.0cm}
       p{2.0cm}
       p{1.5cm}
       p{3.5cm}
       p{4.5cm}
    }
    \hline
    \textbf{Study}
      & \textbf{Architecture }
      & \textbf{Activation}
      & \textbf{Explainability}
      & \textbf{Applications}
      & \textbf{Remarks} \\
    \hline
    STFT \cite{wang2024spatiotemporal}
      & Attention
      & GELU
      & n/a
      & Large‐scale traffic forecasting
      & Using GLU as an intermediate representation layer. \\
    LMHaze \cite{zhang2024lmhaze}
      & SSM
      & Swish$\beta$
      & n/a
      & Imaging dehazing
      & Swish$\beta$ is used in each SSM block sequentially. \\
    Agda \cite{kogkalidis2024learning}
      & Attention
      & Swish$\beta$
      & n/a
      & Proof formalization
      & SwiGLU + residual connections + prenormalization with RMSNorm. \\
    Read-ME \cite{cai2024textit}
      & MoE-Attention
      & Swish$\beta$
      & n/a
      & Resource-constrained LLM inference
      & Inserting a MoE router into the Transformer. \\
    GLUSE \cite{le2025gluse}
      & Attention
      & Sigmoid
      & n/a
      & Onboard satellite EO image classification
      & GLU as a residual block plus SE for adaptive attention. \\
    Adapter \cite{le2024impact}
      & Attention
      & Sigmoid
      & n/a
      & Clinical notes classification
      & Low-Rank Adaptation layers run alongside Transformer layers. \\
    \textbf{This study}
      & \textbf{Attention, GAU}
      & \textbf{11 functions}
      & \textbf{MINE,- t-SNE}
      & \textbf{PICU PPG artifact detection}
      & \textbf{Experiments on activations/attention with explainability.} \\
    \hline
  \end{tabular}
\end{table*}

\begin{algorithm}[!t]
\caption{Mutual Information Neural Estimation (MINE)}
\label{algo:mine}
\footnotesize
\begin{algorithmic}[1]
\State $\theta \leftarrow$ Initialize network parameters
\Repeat
\State Sample a minibatch of $b$ pairs $(x^{(i)}, z^{(i)})$ from the joint distribution $\mathbb{P}_{XZ}$
\State Independently sample $b$ instances $\tilde{z}^{(i)}$ from the marginal distribution $\mathbb{P}_{Z}$
\State Compute the empirical lower bound $V(\theta)$:
\[ V(\theta) \leftarrow \frac{1}{b} \sum_{i=1}^b T_{\theta}(x^{(i)}, z^{(i)}) - \log\left(\frac{1}{b} \sum_{i=1}^b e^{T_{\theta}(x^{(i)}, \tilde{z}^{(i)})}\right) \]
\State Calculate bias-corrected gradient estimates:
\[ G(\theta) \leftarrow \nabla_{\theta} V(\theta) \]
\State Update the network parameters using gradient ascent:
\[ \theta \leftarrow \theta + \alpha G(\theta) \]
\Until{convergence is reached}
\end{algorithmic}
\end{algorithm}
\vspace{-3mm}

\begin{algorithm*}[!htp]
\caption{MINE for a GRN-Transformer}
\label{algo:mine_grn}
\footnotesize
\begin{algorithmic}[1]

\Require{Encoded representations $\textbf{before\_grn}$ and $\textbf{after\_grn}$, Target labels $\textbf{y}$}

\Procedure{MINE}{}
    \Function{InitializeMINEModel}{input\_shape}
        \State Define a neural network with input shape \textit{input\_shape} and output of size 1, representing \( T_{\theta}(X, Y) \)
        \State \Return{Initialized neural network}
    \EndFunction

    \Function{ComputeMI}{data, y, model}
        \State $\textbf{shuffled\_y} \gets \textbf{shuffle(y)}$ \Comment{Shuffle $y$ to create samples from the marginal distribution}
        \State $\textbf{joint\_pred} \gets \textbf{model}([data, y])$ \Comment{Compute predictions for the joint distribution, \( T_{\theta}(X, Y) \)}
        \State $\textbf{marginal\_pred} \gets \textbf{model}([data, \textbf{shuffled\_y}])$ \Comment{Compute predictions for the marginal distribution, \( T_{\theta}(X, \tilde{Y}) \)}
        
        \State Compute empirical mutual information estimate \( MI \) as:
        \[
        MI \gets \frac{1}{b} \sum_{i=1}^b T_{\theta}(x^{(i)}, y^{(i)}) - \log \left( \frac{1}{b} \sum_{i=1}^b e^{T_{\theta}(x^{(i)}, \tilde{y}^{(i)})} \right)
        \]
        where \( b \) is the minibatch size, \( (x^{(i)}, y^{(i)}) \) are joint samples, and \( (x^{(i)}, \tilde{y}^{(i)}) \) are marginal samples.
        
        \State \Return{$MI$}
    \EndFunction

    \State $model\_before\_grn \gets \Call{InitializeMINEModel}{shape\_of(\textbf{before\_grn}[1])}$
    \State $model\_after\_grn \gets \Call{InitializeMINEModel}{shape\_of(\textbf{after\_grn}[1])}$

    \For{$epoch = 1$ to $num\_epochs$}
        \State Train $model\_before\_grn$ on $\textbf{before\_grn}$ and $\textbf{y}$ \Comment{Optimize $T_{\theta}$ to learn dependencies for the representation \textbf{before\_grn}}
        \State Train $model\_after\_grn$ on $\textbf{after\_grn}$ and $\textbf{y}$ \Comment{Optimize $T_{\theta}$ to learn dependencies for the representation \textbf{after\_grn}}
    \EndFor

    \State $MI\_before \gets \Call{ComputeMI}{\textbf{before\_grn}, \textbf{y}, model\_before\_grn}$
    \State $MI\_after \gets \Call{ComputeMI}{\textbf{after\_grn}, \textbf{y}, model\_after\_grn}$

    \State \textbf{Print}("MI before GRN: ", $MI\_before$)
    \State \textbf{Print}("MI after GRN: ", $MI\_after$)

\EndProcedure
\end{algorithmic}
\end{algorithm*}

\subsection{Integrating GRN to Attention Mechanism}
\begin{figure*}[!htp]
	\centering
	\includegraphics[scale=0.4]{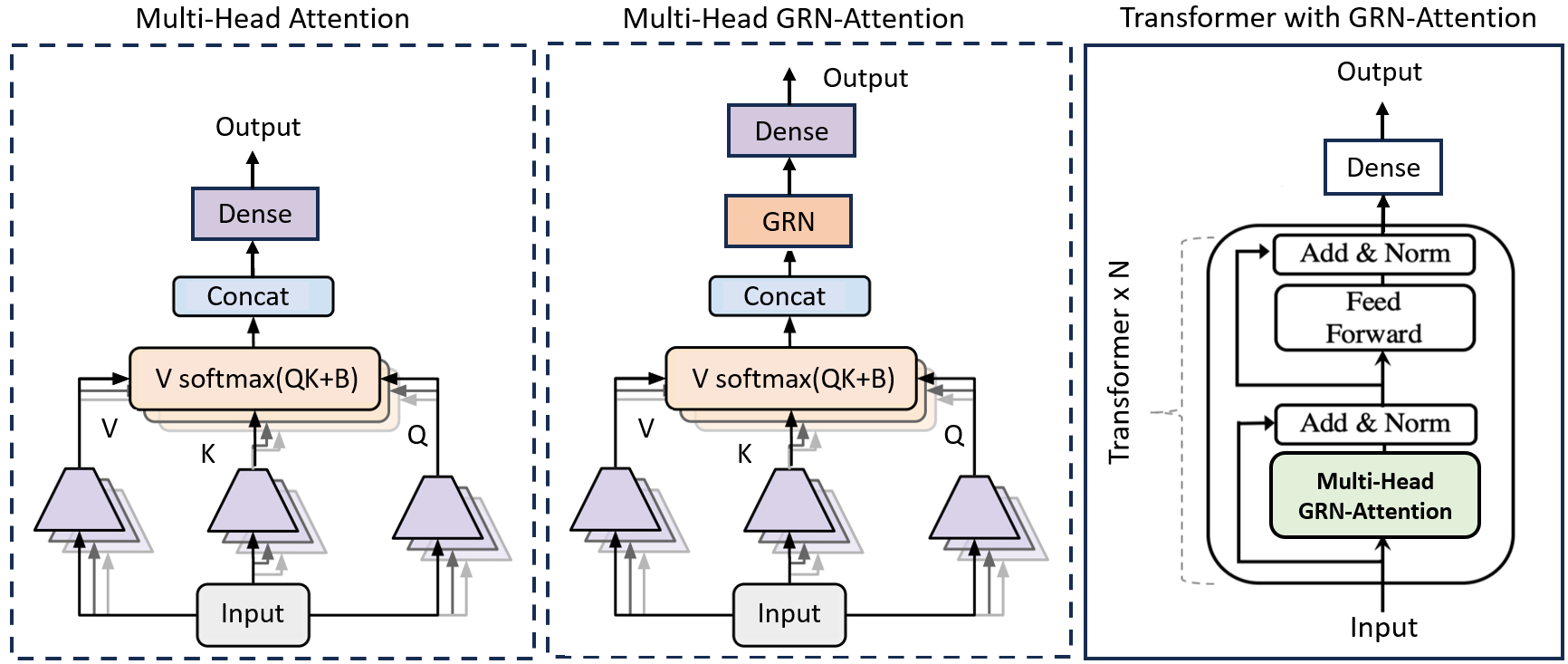}
	\caption{An evolution of the attention mechanism in a Transformer from standard multi-head attention (left) to a GRN-enhanced variant (middle), and the Transformer with GRN-Attention (right).}
	\label{fig:GAU}
\end{figure*}

Inspired by modifications to the attention mechanism proposed in \cite{hua2022transformer}, namely GAU, we investigate integrating the GRN directly into the attention mechanism rather than placing it solely in intermediate layers. This adaptation aims to enhance the model’s ability to capture complex temporal dependencies, potentially improving performance on tasks that rely on time-dependent patterns. Figure \ref{fig:GAU} illustrates the progression of the attention mechanism in our Transformer model, evolving from the original multi-head attention to a GRN-integrated variant. In the standard setup, multi-head attention computes queries, keys, and values through linear transformations, applying softmax to their dot products for attention weighting, followed by concatenation and a dense layer to generate outputs. Our modified approach introduces a GRN between the softmax and concatenation steps, enabling the model to capture sequential dependencies within the data better. The final Transformer model with multi-head GRN-Attention retains the core structure but replaces the conventional attention with the GRN-enhanced variant.

Table \ref{tab:sota_studies} presents a comprehensive comparison of our proposed framework against state-of-the-art studies employing the GRN with Transformer-based approaches, highlighting differences in network structure, activation choices, explainability tools, and target applications. Our work is the first to address physiological-signal PPG artifact detection in the PICU, a notoriously low-data, high-noise setting. It performs a systematic search over 11 activation functions - Linear, Sigmoid, hard\_$\sigma$, SoftSign, Snake, LiSHT, ReLU, ELU, GELU, Swish$\beta$, and Mish—and proves their impact across two distinct network structures, standard Attention and the lightweight GAU. Explainability is built in from the outset by coupling MINE to quantify feature relevance with t-SNE plots that reveal clear class separation. In short, this study does not merely port an existing technique to a new dataset; it introduces architectural, activation-level, and interpretability innovations that unlock Transformer-style models for a previously out-of-reach healthcare use case.


\section{Experimental Results}
\label{sec:result_discussions}

All experiments were conducted on the PICU e-Medical infrastructure, the Miircic Server at CHUSJ. A GPU Quadro RTX 6000 with 24 Gb of memory provided computational capacity for these experiments. The experiments were conducted using the \textit{scikit-learn} library \cite{scikit-learn} and \textit{Keras} \cite{chollet2015keras}. Data was split into 70\% for training and 30\% for testing. Given the complexity of training Transformer models, we focused on four essential hyperparameters—model size, learning rate, batch size, and maximum sequence length—which have been shown to impact the training dynamics of Transformers critically \cite{popel2018training}. Additionally, dropout with a probability of 0.25 \cite{srivastava2014dropout}, the GlorotNormal initializer \cite{glorot2010understanding}, and batch normalization \cite{ioffe2015batch, bjorck2018understanding} were incorporated to enhance model stability. To address class imbalance, we employed the oversampling technique ADASYN \cite{he2008adasyn}. These hyperparameters were fine-tuned to optimize performance while minimizing the risk of overfitting.

\begin{table}[!t]
\centering
\footnotesize
\caption{Hyperparameters of classifiers}
\label{tab:hyperparameters}
\begin{tabular}{|l|l|}
\hline
Hyperparameters                   & Transformer  \\ \hline
Hidden layers                     & 4            \\ \hline
Number of   neurons               & 128          \\ \hline
Number of   multi-heads attention & 4            \\ \hline
Batch size                        & 96           \\ \hline
Dropout                           & 0.25         \\ \hline
Learning rate                     & 6e-04        \\ \hline
Optimizer                         & Adam         \\ \hline
\end{tabular}
\vspace{-3mm}
\end{table}

To effectively assess the performance of our method, metrics including accuracy, precision, recall (or sensitivity), and F1 score. These metrics are defined as follows: 
\vspace{-3mm}

\begin{align}
&\text {Accuracy (acc) }=\frac{\mathrm{TP}+\mathrm{TN}}{\mathrm{TP}+\mathrm{TN}+\mathrm{FP}+\mathrm{FN}} \\ 
&\text {Precision (pre) }=\frac{\mathrm{TP}}{\mathrm{TP}+\mathrm{FP}}  \\
&\text {Recall/Sensitivity (rec)}=\frac{\mathrm{TP}}{\mathrm{TP}+\mathrm{FN}}  \\ 
&\text {F1-Score (f1)} =\frac{2^{\star} \text {Precision}^{\star} \text {Recall}}{\text {Precision }+\text {Recall}} 
\end{align}

\noindent where TN and TP stand for true negative and true positive, respectively, and they are the number of negative and positive patients that are classified correctly. Whereas FP and FN represent false positive and false negative, respectively, and they represent the number of positive and negative patients that were wrongly predicted.

\begin{table}[!htp]
\footnotesize
\centering
\caption{GLU Models' Performance}
\label{table:model_performance}
\begin{tabular}{lcccc}
\hline
Models & Acc $(\uparrow)$ & Pre $(\uparrow)$ & Rec $(\uparrow)$ & F1 $(\uparrow)$ \\ \hline
BilinearGLU & 0.97 & 0.90 & 0.94 & 0.92 \\
GLU & \textbf{0.98} & 0.90 & \textbf{0.97} & \textbf{0.93} \\
hardGLU & 0.97 & 0.89 & 0.93 & 0.91 \\
SoftSignGLU & 0.97 & 0.91 & 0.93 & 0.92 \\
SnakeGLU & 0.97 & 0.91 & 0.92 & 0.92 \\
LiGLU & 0.97 & 0.91 & 0.94 & 0.92 \\
ReGLU & 0.97 & 0.89 & 0.94 & 0.92 \\
EGLU & 0.97 & 0.93 & 0.92 & 0.92 \\
GEGLU & 0.97 & 0.88 & 0.94 & 0.91 \\
SeGLU & 0.97 & 0.90 & 0.93 & 0.92 \\
SwiGLU & 0.97 & 0.91 & 0.93 & 0.92 \\
MiGLU & \textbf{0.98} & \textbf{0.91} & 0.95 & \textbf{0.93} \\
\hline
\end{tabular}
\vspace{1mm} 
\captionsetup{font=scriptsize, justification=raggedright, singlelinecheck=false} 
\caption*{\hspace{5mm}\textbf{Bold} denotes the best values.}
\vspace{-3mm}
\end{table}

Table \ref{table:model_performance} comprehensively compares various Gated Linear Unit (GLU) model variants across four key performance metrics. Among these, the GLU and MiGLU models emerge as the top performers. Both models achieve the highest accuracy, scoring 0.98, indicating their superior overall prediction correctness. The MiGLU model leads in precision, boasting a score of 0.91, which suggests it is particularly effective in making accurate positive predictions with the least false positives. On the other hand, the GLU model excels in recall with a top score of 0.97, highlighting its capability to identify the majority of true positive cases correctly. Furthermore, both these models share the highest F1 score of 0.93, illustrating their optimal balance between precision and recall. The GLU and MiGLU models demonstrate the best overall performance among the variants, making them potentially more effective for tasks requiring high accuracy, precise predictions, and reliable identification of true positives.

\begin{table}[!htp]
\footnotesize
\centering
\caption{GnLU Models' Performance}
\label{table:gnlu_performance}
\begin{tabular}{lcccc}
\hline
Models & Acc $(\uparrow)$ & Pre $(\uparrow)$ & Rec $(\uparrow)$ & F1 $(\uparrow)$\\ \hline
GnLU     & \textbf{0.98} & \textbf{0.91} & \textbf{0.96} & \textbf{0.94} \\
LiGnLU   & 0.97 & 0.91 & 0.92 & 0.91 \\
MiGnLU   & 0.97 & 0.90 & 0.94 & 0.92 \\
SeGnLU   & 0.97 & 0.91 & 0.93 & 0.92 \\
SwiGnLU  & 0.97 & 0.90 & 0.92 & 0.91 \\
\hline
\end{tabular}
\vspace{1mm} 
\captionsetup{font=scriptsize, justification=raggedright, singlelinecheck=false} 
\caption*{\hspace{7.2mm}\textbf{Bold} denotes the best values.}
\vspace{-3mm}
\end{table}

Table \ref{table:gnlu_performance} compares various GnLU model variants, evaluated on key performance metrics. Among these, the standard GnLU model distinctly outperforms its counterparts. It achieves the highest accuracy at 0.98, indicating superior overall prediction correctness. It leads in precision with a score of 0.91, highlighting its effectiveness in making accurate positive predictions with minimal false positives. Additionally, the GnLU model excels in recall with the highest score of 0.96, demonstrating its ability to identify the vast majority of true positive cases correctly. Furthermore, it achieves the top F1 score of 0.94, indicating an optimal balance between precision and recall. This comprehensive performance makes the GnLU model highly effective and reliable in diverse scenarios, particularly when recall and accurate identification of true positives are crucial.

\begin{figure*}[!htp]
	\centering
	\includegraphics[scale=0.525]{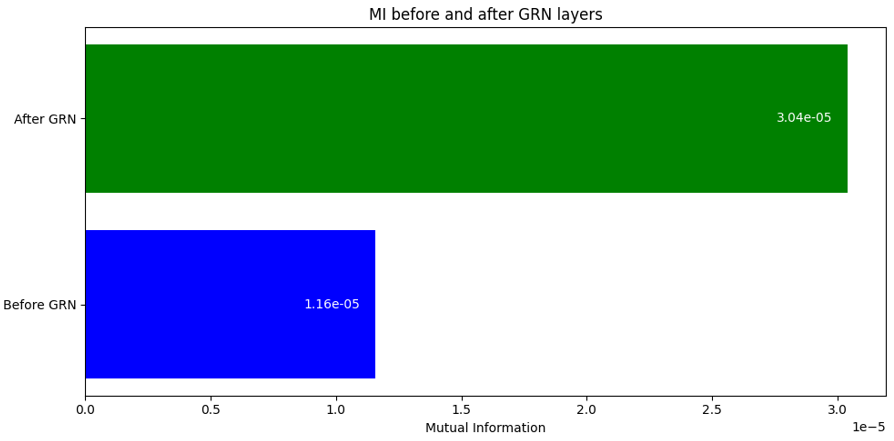}
	\caption{MI Estimation between latent features and class labels before and after the GRN filter block (see Fig. \ref{fig:GRN_mine} for its position in the pipeline). The GRN raises MI, showing that its sigmoid-gated transformation suppresses noise and concentrates task-relevant content, consistent with the subsequent gains in AUC, precision, and recall.}
	\label{fig:mine_estimation}
\end{figure*}

\begin{figure*}[!htp]
	\centering
	\includegraphics[scale=0.5]{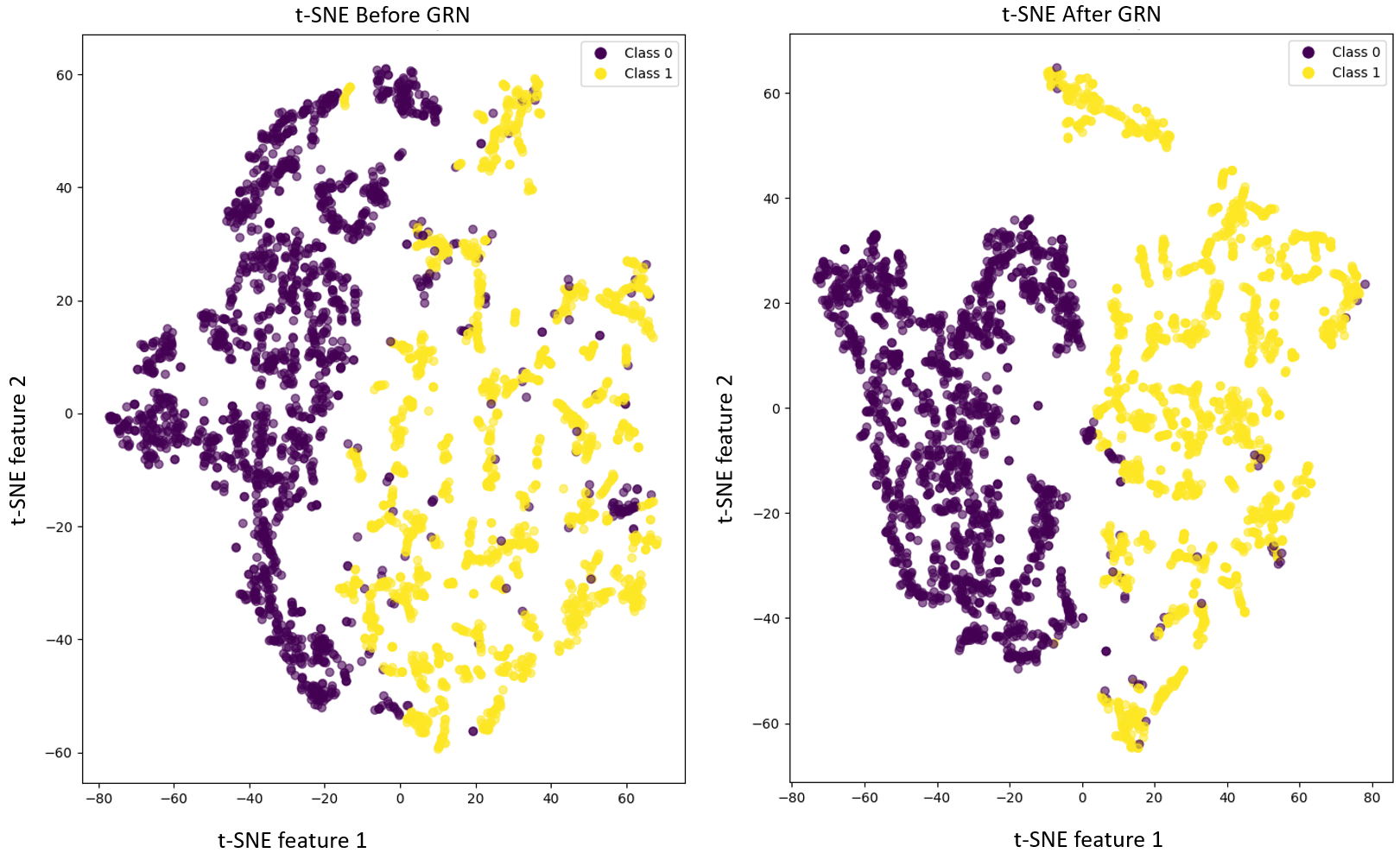}
	\caption{t-SNE visualisation of latent features before (left) and after (right) the Gated Residual Network (GRN) layer. After GRN insertion, the purple (Class 0) and yellow (Class 1) clusters tighten and move farther apart, visibly fewer points straying across the class boundary. This cleaner separation confirms that the GRN transformation filters out noise and concentrates discriminative patterns, mirroring the MI gain in Fig. \ref{fig:mine_estimation}.}
	\label{fig:tsne}
    \vspace{-3mm}
\end{figure*}

\begin{figure*}[!htp]
	\centering
	\includegraphics[scale=0.775]{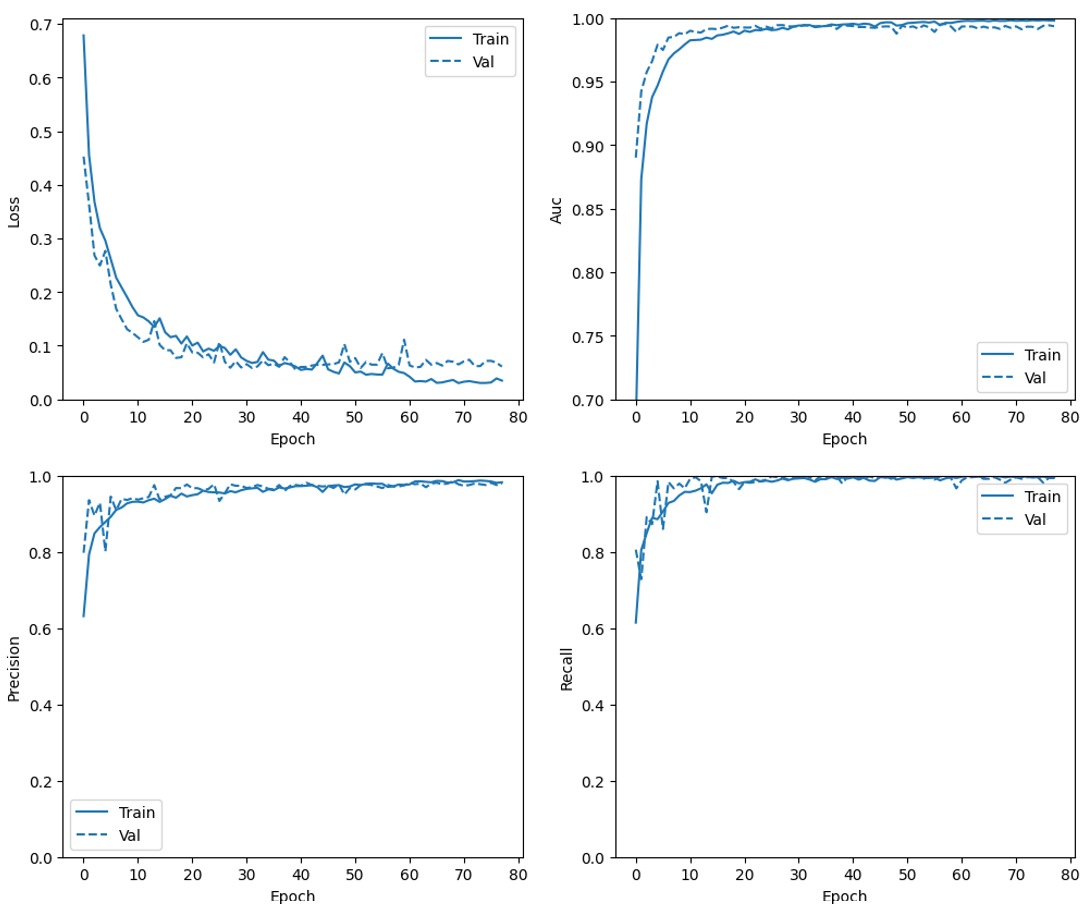}
	\caption{Learning curve during training and validation for the GRN-Transformer that uses a non-linear GLU unit $\sigma(xW + b) \odot \sigma(xV + c)$. The model converges rapidly: validation AUC surpasses 0.98 within ten epochs, while precision and recall stabilise above 0.94 with no divergence between training and validation curves, indicating strong generalisation and performance.}
	\label{fig:learning_gnlu}
    \vspace{-3mm}
\end{figure*}

The provided Fig. \ref{fig:learning_gnlu} depicts the performance of a GRN-Transformer with a GnLU across various training epochs, showing metrics for both training and validation data. The loss graph shows a typical sharp decline in loss for both training and validation at the start of training, which stabilizes as epochs increase, indicating that the model is learning and generalizing well without signs of overfitting. The AUC graph reveals high and stable values for both sets, suggesting excellent class separation capability. Precision and recall start low but rise quickly to plateau at high values, demonstrating that the model accurately identifies true positives and covers a high proportion of actual positive samples. The proximity of training and validation curves across all metrics indicates that the model generalizes well to new data. There's some fluctuation in validation loss, but it's within normal bounds, suggesting some variability in the validation set or learning process. These results point to a well-performing model with strong classification abilities, pending further evaluation in the context of specific project goals and benchmarks.

\begin{figure}[!htp]
	\centering
	\includegraphics[scale=0.4]{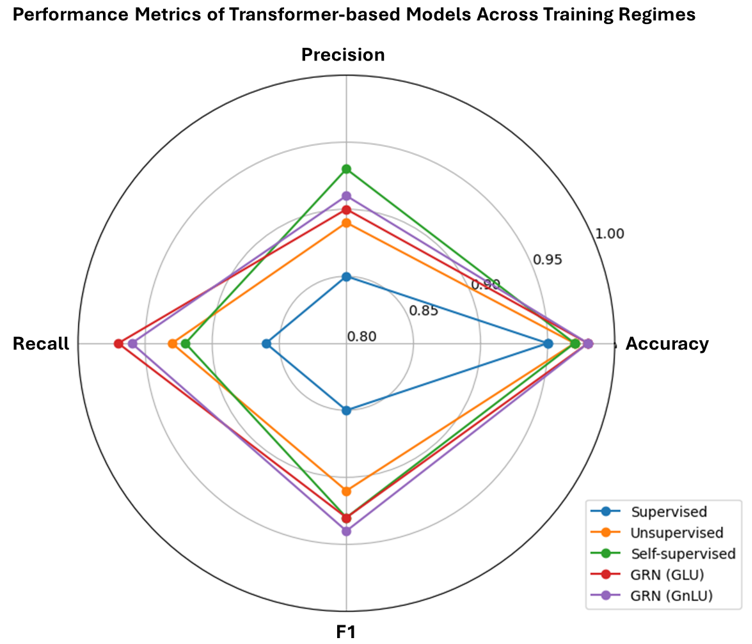}
	\caption{ Performance results from Transformer-based models with different learning paradigms, supervised \cite{macabiau2023label}, unsupervised \cite{le2024boosting}, self-supervised \cite{le2024boosting}, and GRN variants i) with linear GLU unit \cite{le2023grn} ($(xW + b) \odot (xV + c)$), and ii) with non-linear GLU unit ($\sigma(xW + b) \odot \sigma(xV + c)$).}
	\label{fig:chart_perf}
    \vspace{-2mm}
\end{figure}

\begin{figure*}[!htp]
	\centering
	\includegraphics[scale=0.60]{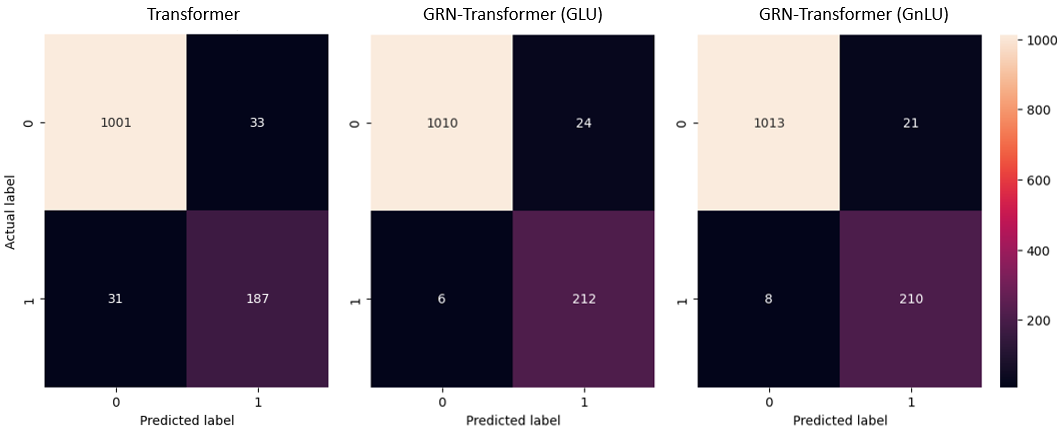}
	\caption{Confusion matrix of the classification task with: \textit{Left}: original Transformer, \textit{Middle}: GRN-Transformer with linear GLU unit ($(xW + b) \odot (xV + c)$), and \textit{Right}: GRN-Transformer with non-linear GLU unit ($\sigma(xW + b) \odot \sigma(xV + c)$).}
	\label{fig:cf_glu}
    \vspace{-5mm}
\end{figure*}

Fig. \ref{fig:chart_perf} shows the comparison of Transformer-based models trained under different paradigms: supervised \cite{macabiau2023label}, unsupervised (AE) \cite{le2024boosting}, self-supervised \cite{le2024boosting}, with GRN \cite{le2023grn}
. Experimental results confirm that both GRN variants, GLU \cite{le2023grn} and GnLN (this study), form the outer‐most polygon, showing the highest recall (0.97 – 0.98) while maintaining near-perfect accuracy (0.98) and strong precision/F1 (0.94). Standard supervised, unsupervised, and self-supervised baselines trail behind, especially on recall and F1. Consequently, GRN-Transformer models combine top-tier accuracy with the best recall, delivering the most balanced F1. Their high recall means far fewer false negatives, an essential property for clinical decision support, where missing a positive case can be costly.  This superiority is further reinforced by the confusion matrix in Fig. \ref{fig:cf_glu}, where GRN-Transformers variants display the lowest false-negative and false-positive cases.

\begin{figure*}[!htp]
	\centering
	\includegraphics[scale=0.75]{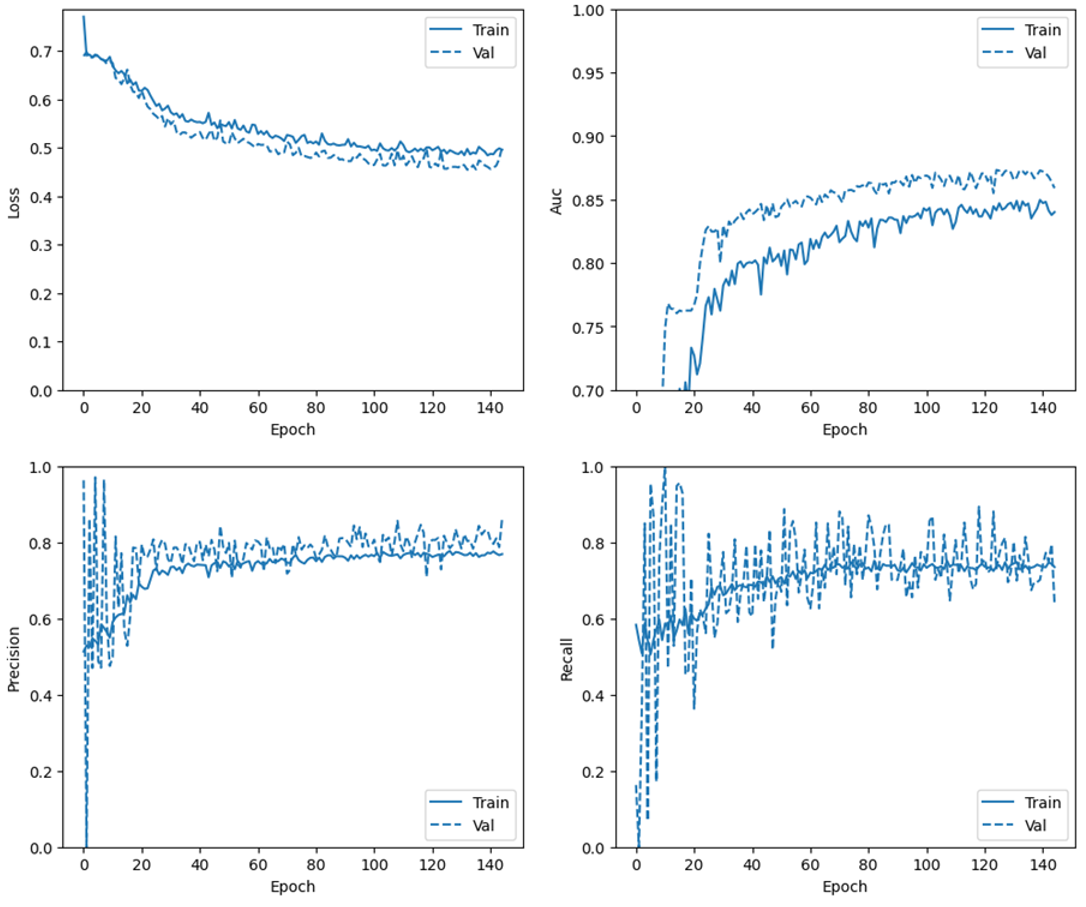}
	\caption{Transformer's learning curve performance with GRN-Attention structure. The higher loss, lower and fluctuating AUC, recall, and precision compared with the conventional Attention (Fig. \ref{fig:learning_gnlu}) indicate that the hybrid GRN-Attention cannot exploit the scarce PICU data as effectively as the conventional Attention design.}
	\label{fig:learning_gau}
    \vspace{-5mm}
\end{figure*}

The Fig. \ref{fig:learning_gau} displays the performance of a Transformer model with GRN-Attention over 140 epochs, showing key metrics for training and validation datasets. The loss graph indicates an initial decrease in training and validation losses, with subsequent fluctuations in validation loss hinting at potential overfitting. The AUC scores begin high and plateau, but a widening gap between training and validation suggests the model may be learning training-specific patterns that don't generalize well. Precision and recall metrics are initially volatile but stabilize, though they exhibit considerable noise, especially in the validation set. This noise implies inconsistency in the model's predictive accuracy and sensitivity to positive samples. While the model shows initial solid learning, the observed metrics suggest it may be overfitting to the training data. Regularization, hyperparameter tuning, or early stopping may be necessary to improve the model's generalization to new data and ensure stable performance across all metrics.

\section{Clinical and Translational Impact}
\label{sec:clinical_implications}

Embedding the proposed GRN-Transformer artifact-detection module into the CHUSJ's CDSS stack will supply bedside teams with cleaner SpO\textsubscript{2} and PPG information, reducing false alarms, minimising manual waveform checks, and accelerating recognition of incipient hypoxaemia or ARDS. When fused with algorithms already validated at CHUSJ—including occlusion-segmentation of monitor traces \cite{munoz2024hybrid}, heart-rate/temperature coupling analysis \cite{lu2025heart}, absence-of-heart-failure extraction from clinical notes \cite{le2022detecting,le2023adaptation,le2023small,lompo2024numerical,lompo2024multi}, hypoxaemia prediction \cite{sauthier2021estimated}, and automated chest x-ray \cite{zaglam2014computer,yahyatabar2020dense}, the system can deliver an integrated, continuously updated cardiorespiratory score that helps clinicians adjust ventilatory settings or order confirmatory tests earlier and with greater confidence.

Deploying the composite CDSS in a clinical setting entails several operational challenges. First, the system must process 128 Hz waveform streams in real time so that alerts are not delayed. Second, it must interface reliably with diverse bedside monitors and the existing HRDB. Third, robust model governance, encompassing periodic retraining, performance audits, and version control, is essential to prevent drift as hardware or patient populations evolve. Fourth, all data flows must comply with institutional and provincial regulations on cybersecurity and patient privacy. Finally, the user interface should minimise alarm fatigue and integrate smoothly with established PICU workflows as proven \cite{yakob2024data, lu2025heart}. This evaluation will guide further system refinements and improvements to quantify latency, false-alarm reduction, and clinician acceptance before the system becomes a routine part of PICU care.

\section{Conclusion}
\label{sec:conclusion}
This study explores the Transformer model's performance in various learning environments, focusing on GRN-Transformer. Our research analyzes different activation functions for the GLU, a crucial component of the GRN structure. The study employs MINE to verify the effectiveness of GRN. Additionally, it investigates the positional impact of GRN within the Transformer architecture, particularly examining its role as an intermediate layer within the Attention mechanism instead of an external intermediary layer. Results show that the GnLU with Sigmoid gating consistently yields the best accuracy, precision, recall, and AUC; MINE confirms the GRN’s insertion increases feature-label MI; and positioning the GRN as an intermediate representation filter layer - rather than embedding it inside the Attention block - more effectively suppresses noise and amplifies clinically relevant patterns.

This work explicates how a GRN acts as an intermediate‐representation filter, suppressing noise, amplifying salient features, and measurably increasing feature-label MI to boost Transformer performance when labelled data are scarce.  By hybridizing this architectural unit with a systematic activation analysis and an MI-based interpretability pipeline, we deliver methodological advances that extend beyond the PICU use case: the GRN block is model-agnostic, computationally lightweight, and can be dropped into any encoder–decoder stack or temporal-sequence model where high noise and low annotation density prevail (e.g., wearable sensing, industrial IoT, or low-resource speech).  These findings, therefore, contribute to core neural-network design and analysis while simultaneously offering a practical pathway for deploying models in challenging clinical and other real-world environments.

\section*{Acknowledgment}

The Research Center at CHU Sainte-Justine Hospital, University of Montreal, provided the clinical PPG data. The authors thank Clara Macabiau and Dr. Kevin Albert for their support in data preprocessing and annotating for this research. Data and reproducible codes are available upon reasonable request to Prof. Philippe Jouvet, M.D., PhD. (Email: philippe.jouvet.med@ssss.gouv.qc.ca).

\bibliographystyle{IEEEtran}
\bibliography{IEEEabrv,Bibliography}

\end{document}